\documentclass[longbib,twocolumn,twocolappendix]{aastex701} 
\usepackage{amsmath}
\usepackage{CJK}

\graphicspath{{./}{fig/}}

\shorttitle{Covertly Active Comet (139359) 2001~ME$_1$}
\shortauthors{Q. Zhang et al.}

\journalinfo{The Planetary Science Journal, in press} 

\received{2025 June 20}
\revised{2025 November 25}
\accepted{2025 December 22}

\begin{document}
\begin{CJK*}{UTF8}{gbsn}

\title{Covertly Active Comet (139359) 2001~ME$_1$}

\correspondingauthor{Qicheng Zhang}

\author[0000-0002-6702-191X]{Qicheng Zhang}
\affiliation{Lowell Observatory, 1400 W Mars Hill Rd, Flagstaff, AZ 86001, USA}
\email[show]{qicheng@cometary.org}

\author[0000-0002-4838-7676]{Quanzhi Ye (叶泉志)}
\affiliation{Department of Astronomy, University of Maryland, College Park, MD 20742, USA}
\affiliation{Center for Space Physics, Boston University, 725 Commonwealth Ave, Boston, MA 02215, USA}
\email{}

\author[0000-0002-8692-6925]{Karl Battams}
\affiliation{US Naval Research Laboratory, 4555 Overlook Ave, SW, Washington, DC 20375, USA}
\email{}

\author[0000-0003-2781-6897]{Matthew M. Knight}
\affiliation{Physics Department, United States Naval Academy, Annapolis, MD 21402, USA}
\email{}

\author{Worachate Boonplod}
\affiliation{The Thai Astronomical Society, Bangkok, Thailand}
\email{}

\author{Rainer Kracht}
\affiliation{Ostlandring 53, D-25335 Elmshorn, Schleswig-Holstein, Germany}
\email{}

\begin{abstract}
On 2018~November~18, coronagraphs onboard the Solar and Heliospheric Observatory (SOHO) captured an unrecognized comet crossing its fields of view. We identified this comet to be the minor planet (139359) 2001~ME$_1$ whose previously unnoticed dust activity near perihelion became optically amplified by efficient forward scattering of sunlight as the comet crossed between the Sun and SOHO/Earth at up to $175^\circ\llap{.}6$ phase angle. Simultaneous backscattering observations by the Solar Terrestrial Relations Observatory (STEREO) precisely constrain the comet's $\sim$7~mag forward scattering brightening, enabling a direct comparison with the $\sim$3~mag brightening of the more active but optically dust-poor comet 2P/Encke seen by SOHO and STEREO under similar geometry in 2017. Earlier STEREO observations from 2014 additionally show the newly recognized activity to be recurrent---consistent with a reanalysis of the comet's associated meteor activity---and has likely only been previously overlooked due to the comet's faintness and proximity to the Sun while active. Orbital integrations show the comet has likely followed a near-Earth orbit for at least the past 10~kyr, suggesting that the weakness of its observed activity evolved through its continued depletion of accessible volatiles.
\end{abstract}

\keywords{\uat{Comets}{280} --- \uat{Meteor streams}{1035} --- \uat{Near-Earth objects}{1092} --- \uat{Short period comets}{1452} --- \uat{Coma dust}{2159} --- \uat{Comet dynamics}{2213} --- \uat{Comet dust tails}{2312}}

\section{Introduction}

The population of Near-Earth objects (NEOs) is largely derived from asteroids from the main belt interspersed with comets from the scattered disk of the Kuiper Belt \citep{bottke2002}. The asteroid--comet distinction has become blurred in recent years, with the discovery of otherwise unremarkable main-belt asteroids exhibiting comet-like activity driven by water ice sublimation \citep{hsieh2006,kelley2023} alongside the growing number of apparently inactive minor planets with orbits and surface properties resembling comet nuclei from the outer solar system---likely the dormant nuclei of formerly active comets whose past activity have exhausted their accessible supply of near-surface ices \citep{fernandez2002}. A few of these latter objects have moreover been found to still exhibit low levels of residual activity upon close inspection \citep{mommert2014,ye2016b,li2017}.

Minor planet (139359) 2001~ME$_1$ is one NEO previously suspected of being a dormant comet nucleus, with a $\sim$3\% geometric albedo and X-class optical/near-infrared surface colors common for comet nuclei \citep{rivkin2005,mommert2020,masiero2021}. Its orbit, with a Tisserand's parameter of 2.7 with respect to Jupiter, is also more typical of Jupiter-family comets from the outer solar system than asteroids from the main belt \citep{levison1997}, although the current perihelion distance $q=0.34$~au is smaller than that of most NEOs of any variety. Additionally, the Canadian Meteor Orbit Radar \citep[CMOR;][]{jones2005} clearly detected an outburst of meteors in 2006 attributed to this object, indicating it must have been actively producing dust within the prior century \citep{ye2016a}. Furthermore, the astrometry of (139359) indicates the object exhibits comet-like nongravitational acceleration above that expected from Yarkovsky drift, indicative of unseen outgassing \citep{seligman2024}. However, no direct detections of gas or dust production have previously been reported, even with dedicated ground-based activity searches \citep{mommert2020,seligman2024}.

Consequently, when (139359) appeared as a prominently active comet in images from the Large Angle and Spectrometric Coronagraph \citep[LASCO;][]{brueckner1995} cameras onboard the Solar and Heliospheric Observatory \citep[SOHO;][]{domingo1995}, its identity was not immediately recognized. Over its lifetime, LASCO has recorded several thousand new comets near the Sun unseen by any other telescopes, most of which were found and reported by citizen scientists to the Sungrazer project, which curates these discoveries \citep{battams2017}. On 2018 November 18, co-author W. Boonplod reported what appeared to be another one of these comets crossing the LASCO fields of view. We subsequently located this comet in imagery from the Heliospheric Imager 1 \citep[HI1;][]{eyles2009} instrument onboard a Solar Terrestrial Relations Observatory \citep[STEREO;][]{kaiser2008} spacecraft, and realized after performing an orbital fit to the astrometric positions measured from both datasets that this ``new'' comet---which had been provisionally designated SOHO-3651---was actually (139359).

In the following sections, we present and analyze these and other observations of (139359)'s cometary activity. We discuss the forward scattering brightening of the comet's dust seen by SOHO, which we use to characterize the comet's dust properties for comparison with those of 2P/Encke---a brighter and better observed comet with comparable $q$ previously seen under similar observing geometry. We then evaluate the dynamical history of the comet to place its modern day activity in the context of its thermal and physical evolution, and use the observed cometary activity to characterize the comet's associated meteor activity.

\section{Data}

\subsection{Instruments}

\begin{figure*}
\includegraphics[width=\linewidth]{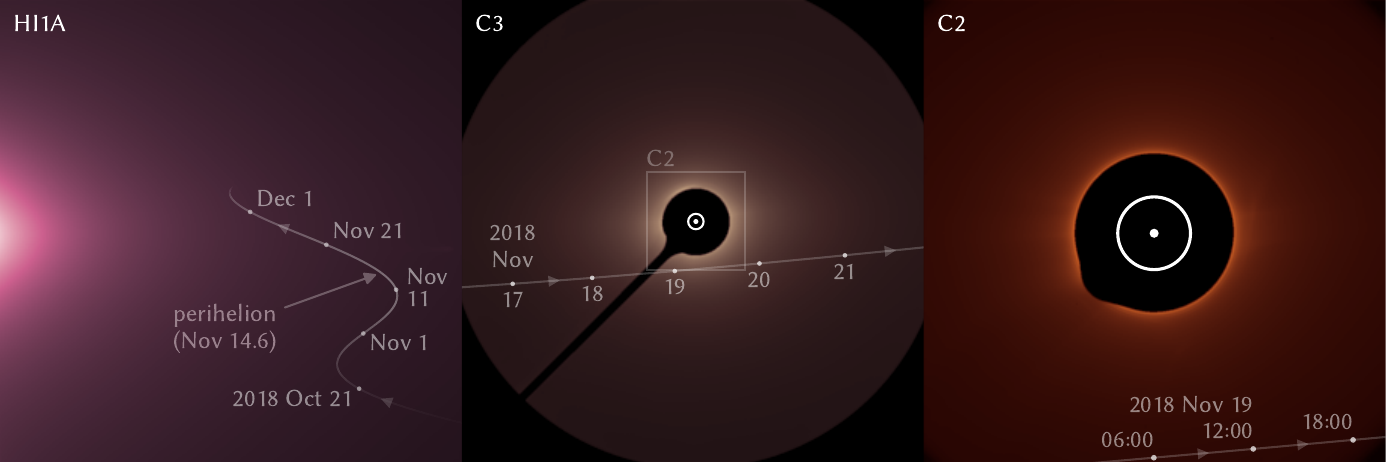}
\caption{Trajectory of (139359) 2001~ME$_1$ during its 2018 apparition across the STEREO-A HI1 (HI1A; left) and SOHO LASCO C3/C2 (center/right) fields of view. The size and position of the Sun behind the LASCO occulters are indicated by the dotted circles ($\odot$). The Sun is off the left edge of the HI1 frame.}
\label{fig:traject}
\end{figure*}

\subsubsection{SOHO LASCO C2/C3}

The SOHO spacecraft monitors the heliospheric space environment from the vicinity of Earth with its two LASCO cameras \citep{brueckner1995}. The inner camera, LASCO C2, monitors the narrow region 1.5--6~$R_\odot$ above the solar limb at 12~arcsec~px$^{-1}$, normally through . The outer camera, LASCO C3, covers a wider region 3.7--30~$R_\odot$ from the limb at a coarser 56~arcsec~px$^{-1}$, normally through a $\sim$530--840~nm FWHM clear filter. Under the synoptic observing sequence that runs by default, full resolution frames are only captured through these standard filters. While images can be taken through several additional color filters with special observing sequences planned in advance, we only have synoptic observation data available for our present analysis due the serendipitous nature of the comet's detection.

We reduced the publicly available level~0.5 data in a similar manner to \citet{zhang2023}, and applied bias and vignetting corrections as in the standard level~1 reduction \citep{thernisien2003}, but without full frame distortion/warp correction in order to preserve pixel-scale detail. We then used the provided distortion coefficients and spacecraft pointing to produce an astrometric solution for each frame, corrected through fitting stars from Gaia DR3 \citep{gaia2023}, and extracted subsampled, comet-centered subframes using ephemerides from JPL Horizons and JPL orbit \#130. We stacked the resulting comet-centered subframes with $2.5\sigma$ rejection of solar energetic particles, cosmic rays, stars, and other image defects to improve sensitivity.

\subsubsection{STEREO-A HI1}

The STEREO mission is comprised of two spacecraft, with one, STEREO Ahead (STEREO-A), orbiting the Sun slightly interior to and thus faster than Earth, and the other, STEREO Behind (STEREO-B), orbiting slightly exterior to and thus slower than Earth \citep{kaiser2008}. Each spacecraft has an effectively identical suite of instruments, of which, we used only data from the Heliospheric Imager 1 (HI1) of the Sun Earth Connection Coronal and Heliospheric Investigation \citep[SECCHI;][]{howard2008} instrument suite. Each HI1 camera captures a $20^\circ\times20^\circ$ field at 72~arcsec~px$^{-1}$ aimed along the Sun--Earth line, covering the portion between $4^\circ$ and $24^\circ$ solar elongation. Both HI1 cameras observe through a fixed bandpass filter with a preflight FWHM span of $\sim$615--740~nm, a blue leak near 400~nm, and a red leak near 1000~nm \citep{bewsher2010}, although more recent analysis suggests its bandpass may have shifted $\sim$20~nm blueward from exposure to the space environment \citep{halain2012}. STEREO-B failed in 2014, leaving only STEREO-A in operation afterward \citep{ossing2018}, and we only use data from STEREO-A in our present analysis.

We used the publicly available level~2 data for our analysis, which are bias- and vignetting-corrected, astrometrically-calibrated, and background-subtracted to remove the F-corona gradient. We used the level~2 images with F-corona backgrounds generated as the running mean of the lowest quartile frames over 11~day windows. As with the LASCO data, we extracted subsampled, comet-centered subframes using JPL Horizons ephemerides. However, prior to stacking, we also subtracted a siderally-aligned stack of surrounding frames to mitigate the impact of the dense stellar background at the high sensitivity and low resolution of HI1.

\subsection{Observations}

\begin{figure*}
\centering
\includegraphics[width=\linewidth]{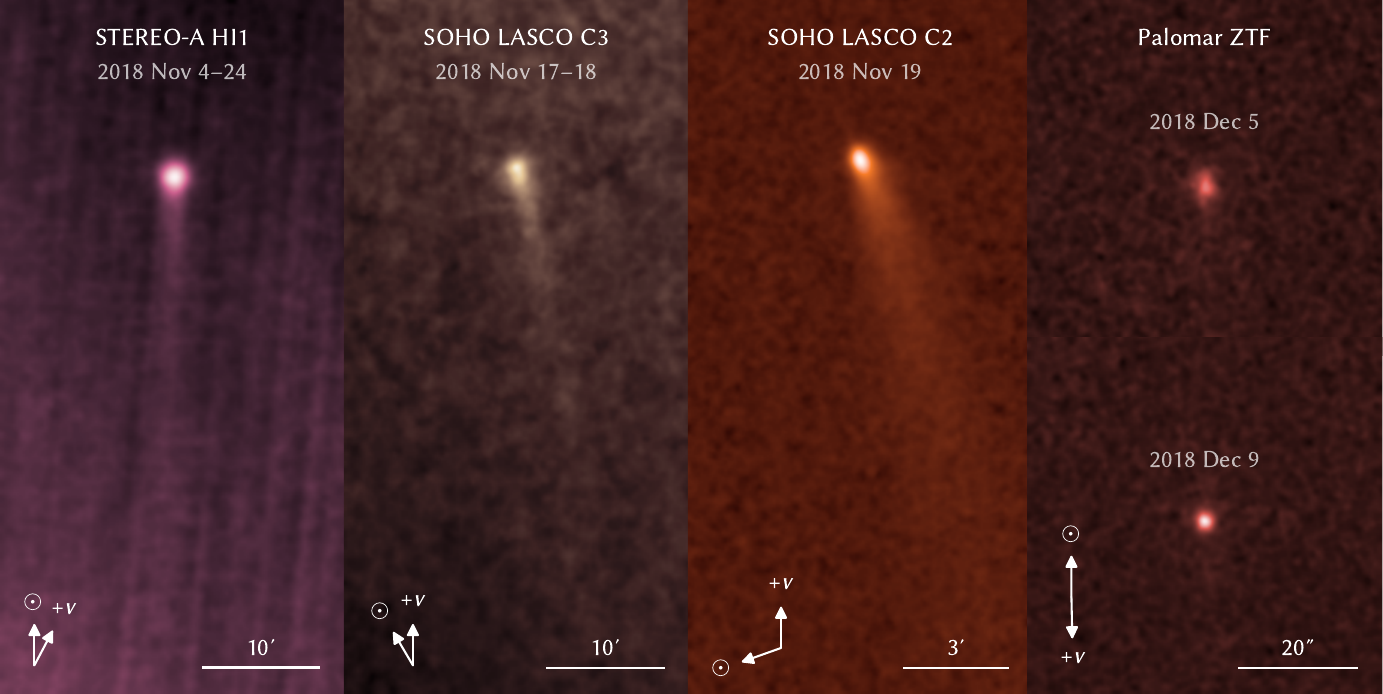}
\caption{Stacked frames of (139359) 2001~ME$_1$ showing its tail during its 2018 apparition from STEREO-A HI1, SOHO LASCO C3 (clear-filtered) and C2 (orange-filtered), and Palomar ZTF ($r'$-filtered). Prior to stacking, the constituent HI1 and ZTF frames were rotated to place the sunward direction ($\odot$) upward, while those of the LASCO stacks were rotated to place the direction of the comet's heliocentric velocity ($+v$) upward. In the ZTF images, the comet appears slightly diffuse with a hint of an antisunward tail on December~5 but becomes indistinguishable from a point source by December~9 as the comet moved away from the Sun ($r=0.63$~au to 0.70~au) and the nucleus brightened relative to the dust with the decreasing phase angle ($\alpha=98^\circ$ to $89^\circ$) over this period.}
\label{fig:img}
\end{figure*}

During its 2018 apparition, (139359) reached perihelion on November~14.9 and subsequently crossed between the Sun and the Earth, reaching a maximum phase angle $\alpha=175^\circ\llap{.}6$ (scattering angle $\theta=180^\circ-\alpha=4^\circ\llap{.}4$) near inferior conjunction at November~19.3. Both the LASCO cameras captured this crossing, aided by the dust's efficient forward scattering of sunlight \citep{marcus2007}, as discussed later in section~\ref{sec:forward}. The comet crossed the field of the LASCO C3 from November~16 through 21, and passed within that of LASCO C2 on November~19. The comet also crossed the STEREO-A HI1 field over several months around this period at much lower $\alpha\sim60^\circ$ ($\theta\sim120^\circ$), and was sufficiently bright to be visible in 2~day stacks through most of 2018~November. Figure~\ref{fig:traject} shows the trajectory of the comet across the fields of view of all three cameras in this apparition.

We also note the Zwicky Transient Facility \citep[ZTF;][]{bellm2019} observed the comet later this same apparition. ZTF is a wide-field sky survey using the 1.2~m Samuel Oschin Telescope at Palomar Observatory to target astronomical transients in the night sky \citep{graham2019}, but also conducts twilight observations targeting solar system objects near the Sun \citep{ye2020}. A stack of $r'$-filtered frames from twilight observations on 2018~December~5 UT captured the comet receding from perihelion at $r=0.63$~au as a slightly diffuse source with a trace of an antisunward tail \citep{irsa2022}. This diffuseness could reflect the comet's residual activity or the presence of slow-moving dust grains ejected earlier near perihelion remaining near the nucleus. That morphology rapidly transitioned to become nearly stellar by December~9 UT, likely due to both the comet's decreased activity at $r=0.70$~au and lower $\alpha=89^\circ$ (compared to $98^\circ$ on December~5) brightening the nucleus relative to the surrounding dust, since the solid nucleus monotonically brightens with decreasing $\alpha$ at all $\alpha$ while side-scattering cometary dust grains dim with decreasing $\alpha$ in this range \citep{schleicher2011}. ZTF twilight observations in 2023 at $r=0.65$~au post-perihelion likewise showed the comet with similarly extended morphology. We did not use the ZTF data apart from this morphology evaluation.

Figure~\ref{fig:img} shows the comet's morphology during the 2018 apparition in stacks of frames from all of these instruments. The STEREO-A HI1 and SOHO LASCO stacks both captured the comet near perihelion with a prominent tail visibly extending ${>}30'$ in the HI1 and C3 stacks, while the December~5 ZTF stack shows a more limited tail only a few arcseconds long a few weeks later.

We also located the comet near perihelion in STEREO-A HI1 imagery during its 2014 apparition when it exhibited similar behavior as 2018, and we incorporated that data into our photometric analysis with the 2018 apparition SOHO/STEREO data in section~\ref{sec:phot}. The comet also previously crossed the SOHO LASCO C3 field in 2001 at up to $\alpha=171^\circ\llap{.}4$ while pre-perihelion at $r=0.64$~au, but was not definitively detected above the noise level in a 4~day stack.

\section{Analysis}

\subsection{Photometry}
\label{sec:phot}

\begin{figure*}
\centering
\includegraphics[width=0.495\linewidth]{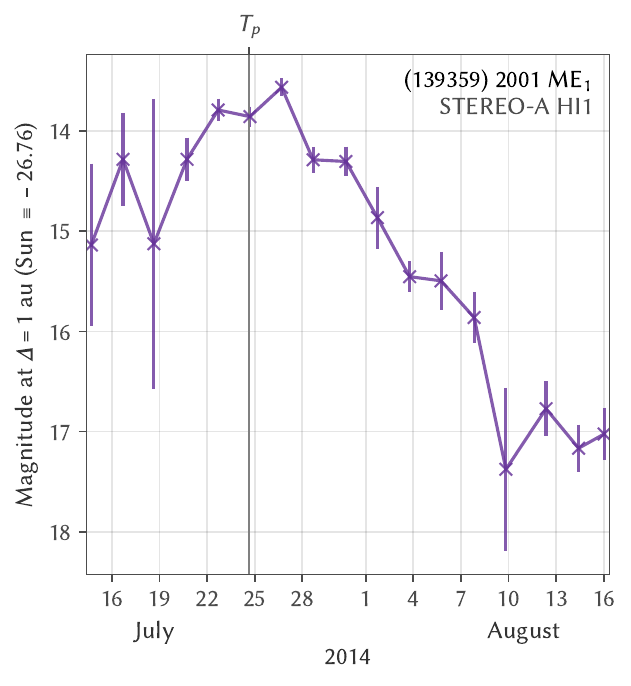}
\includegraphics[width=0.49\linewidth]{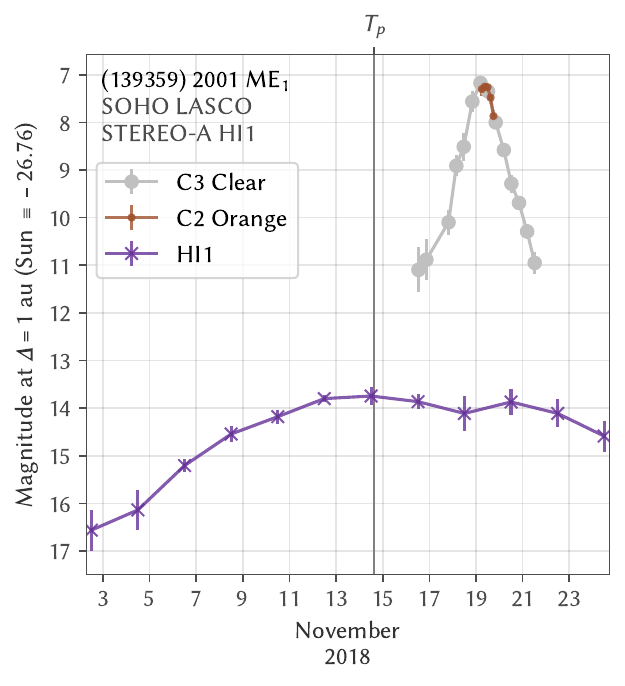}
\caption{Photometry of (139359) 2001~ME$_1$ measured from SOHO LASCO C2/C3 and STEREO-A HI1 imagery in 2014 (left) and 2018 (right) with $3'$ radii apertures, normalized to $\varDelta=1$~au. The times of perihelion are indicated by $T_p$. During its 2018 apparition when the comet was simultaneously observed by both spacecraft, the comet appeared much brighter from SOHO than from STEREO due to the much higher $\alpha$ of the comet from SOHO, demonstrating the high efficiency of forward scattering by cometary dust.}
\label{fig:phot}
\end{figure*}

We used only the SOHO and STEREO data from the 2014 and 2018 apparitions, which captured the comet near perihelion, for our photometric analysis. We measured the brightness of the comet in image stacks of consecutive frames over 3~hr intervals for LASCO C2, 8~hr intervals for LASCO C3, and 2~day intervals for HI1. For all data, we performed photometry with circular apertures with $3'$ radius, using the LASCO photometric calibration from \citet{zhang2023} and the HI1 photometric calibration from \citet{tappin2022}. This aperture size is around the smallest that avoids aperture losses above the $\sim$0.1~mag level in the lowest resolution, HI1 imagery. We estimated the photometric uncertainty from the spatial variation in the background surrounding the photometric aperture.

We used the standard solar magnitude system, where the Sun from $r=1$~au away is defined to be of magnitude $-26.76$ (i.e., equal to its Johnson $V$ magnitude from $r=1$~au) in every bandpass. Figure~\ref{fig:phot} shows the resulting light curves from both the 2014 and 2018 apparitions.

Finally, for our analysis of the comet's dust and gas activity, we subtracted the nucleus contribution to the observed flux, modeled as a linear phase function with an absolute magnitude of 16.6 and a slope of 0.043~mag~deg$^{-1}$, as fitted to observations published by the Minor Planet Center through Minor Planet Circulars Supplement 2161850. As the nucleus was fainter than the coma in all of our photometric data---$<$0.1\% of the forward scattering coma brightness in its 2018 LASCO transit and still only $\sim$10\% of the backscattering brightness in HI1 during the same period---this correction had a minor impact on the photometry. Errors in the assumed nucleus phase function therefore had negligible impact the results, especially at higher $\alpha$ where the linearly extrapolated nucleus phase function likely has a larger relative error, but becomes dwarfed by an enormous brightening of the dust in the coma, as discussed in section~\ref{sec:forward}.

\subsubsection{Forward Scattering}
\label{sec:forward}

Comets crossing between the Sun and Earth at high $\alpha$ (low $\theta$) often appear to undergo a temporary surge in brightness attributed to the high forward scattering efficiency of typical micron-sized dust grains, a diffraction effect responsible for producing many of the brightest comets in history \citep{marcus2007}. The diffraction pattern of a dust grain is strongly dependent on its physical size, so the shape of the brightness surge provides an avenue to measure the size distribution of the coma dust population. Additionally, as gaseous species emit in a far more isotropic manner than dust, the strength of a comet's forward scattering brightness surge provides a useful constraint on the relative abundance of optically visible dust and gas in the coma in the absence of spectra or multicolor imaging.

However, while comets routinely pass through forward scattering geometry, most do so unseen from Earth, as comets at very high $\alpha$ must necessarily be in close angular proximity to the Sun where all but the brightest comets become drowned out by twilight or daylight. The sunward-pointing imagers of the SOHO and STEREO spacecraft, however, are well-suited to capturing comets brightened by forward scattering. Several such comets have now been seen by both missions reaching $\alpha>170^\circ$ ($\theta<10^\circ$) where the corresponding brightening surge can reach several orders of magnitudes in amplitude \citep{hui2013,knight2024}.

\begin{figure}
\includegraphics[width=\linewidth]{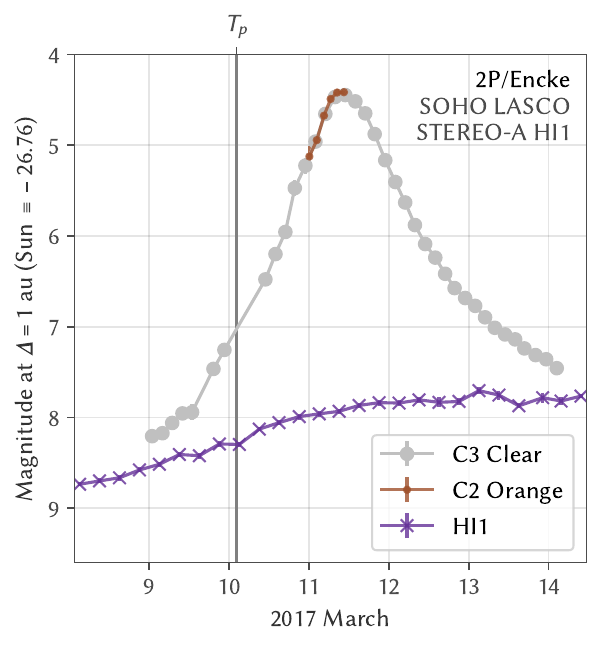}
\caption{Forward scattering brightness surge of 2P/Encke seen by SOHO LASCO C2/C3 with simultaneous baseline photometry from STEREO-A HI1 during the comet's 2017 apparition, with $T_p$ indicating the time of perihelion.}
\label{fig:lc_2P}
\end{figure}

Even so, (139359) serves as an unusual case of a comet simultaneously observed at both high and low $\alpha$. Such data is particularly useful in disentangling the forward scattering brightening from temporal variations in the comet's activity, enabling precise constraints on the former. The only other instance of a known comet simultaneously seen at high $\alpha>170^\circ$ by SOHO and at low $\alpha<90^\circ$ by STEREO is 2P/Encke during its 2017 apparition. Shortly after its March~10 perihelion, 2P crossed through both the SOHO LASCO C3 and C2 fields of view, reaching a peak of $\alpha=175^\circ\llap{.}0$ ($\theta=5^\circ\llap{.}0$) on March~11, while simultaneously crossing the STEREO-A HI1 field at much lower $\alpha\approx30^\circ$ ($\theta\approx150^\circ$). We likewise measured the brightness of 2P through all cameras as done for (139359), but with narrower time windows for stacking of 6~hr for HI1 and 3~hr for C3 due to 2P being much brighter. Figure~\ref{fig:lc_2P} shows the resulting photometry of 2P, illustrating its forward scattering brightness surge seen from SOHO, which is noticeably weaker than that seen for (139359) at similar $\alpha$.

\begin{figure}
\includegraphics[width=\linewidth]{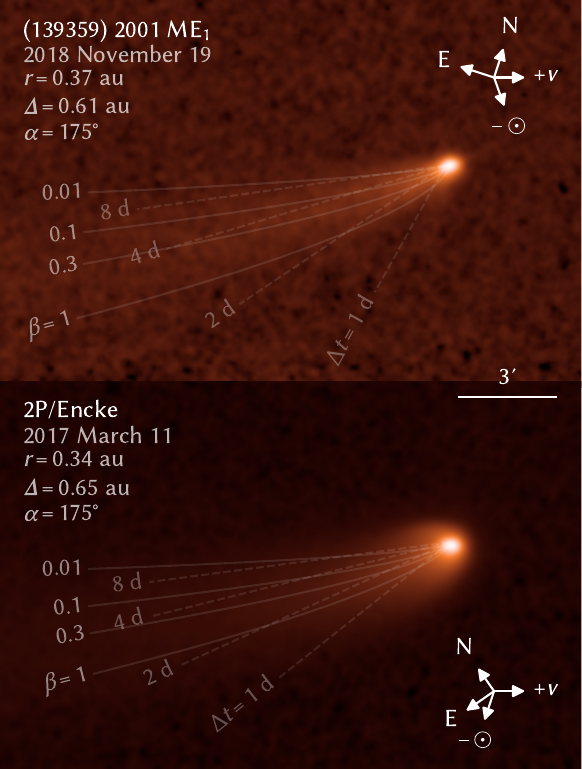}
\caption{Comparison of comets (139359) 2001~ME$_1$ (top) with 2P/Encke (bottom) as seen by SOHO LASCO C2 under similar forward scattering geometry near $\alpha=175^\circ$. Syndyne curves (solid) corresponding to dust of $\beta=1$, 0.3, 0.1, and 0.01, and synchrone curves (dotted) corresponding to ages $\Delta t=1$, 2, 4, and 8~days are overlaid. Note the larger coma of 2P/Encke, which reflects the higher dust ejection velocity supported by its stronger outgassing.}
\label{fig:syn}
\end{figure}

Figure~\ref{fig:syn} compares the morphology of (139359) and 2P near maximum $\alpha$ as seen by SOHO LASCO C2 shortly after perihelion under nearly identical viewing geometry. In addition to being much brighter than (139359), 2P has a noticeably broader, resolved coma, likely from its higher gas production rate accelerating dust away from the nucleus to much higher speeds. Our $3'$ radius photometric aperture readily covers the coma of both comets, but also a variable length of tail that differs not only in time as the comet passes inferior conjunction, but also between SOHO and STEREO due to their vastly different distance and viewing geometry. A proper comparison of dust brightness across $\alpha$ must therefore correct for the different lengths of tail captured by the $3'$ photometric aperture from the two spacecraft.

The curvature of a dust tail for any given comet is largely determined by the predominant size of the constituent dust grains. Smaller grains have larger cross section-to-mass ratios that cause them to accelerate antisunward by solar radiation pressure more quickly, and thus follow trajectories less curved by the comet's orbital motion around the Sun than do larger grains. This property is commonly parametrized for each grain size by the ratio $\beta$ of radiation-to-gravitational force acting on that grain, which is approximately $\beta\sim0.6~\mathrm{\mu m}/r_d$ for an absorbing dust grain of radius $r_d\gtrsim1$~$\mu$m with $\rho_d\sim1$~g~cm$^{-3}$ density at all $r$, since solar radiation and gravity both scale with $r^{-2}$ \citep{finson1968,gustafson2001}.

Syndyne curves, representing dust of constant $\beta$ released from the nucleus at rest, show micron-scale $\beta\sim0.1$--0.3 dust to be optically dominant in both comets in their respective forward scattering geometries, with the spine of (139359)'s tail falling closer to the lower end and that of 2P closer to the upper end. We also plotted synchrone curves, corresponding to dust of constant age $\Delta t$, which show the visible dust of both comets to be largely a few days old; some age modulation of the dust tail brightness may also be evident for (139359), which passed perihelion 4~days earlier, around when it likely had higher dust production activity. We approximated the dust tails of each comet as narrow tubes following the average syndynes, determined the age of the dust where the respective syndynes exit the photometric aperture, and assumed the amount of dust within the aperture is proportional to that age. We applied this assumption to correct the SOHO LASCO photometry to provide the brightness of the same portion of dust tail captured by the simultaneous STEREO-A HI1 photometry. Note, however, that due to the concentration of light within the aperture toward the coma, this correction actually had little effect on the results, with order-of-magnitude variations in the selected $\beta$ corresponding to only a few 0.1~mag of difference in the relative photometry.

The Henyey--Greenstein phase function \citep{henyey1941}---an empirical model approximating the scattering of interstellar dust grains---is commonly used to model the forward scattering brightness of cometary dust \citep{marcus2007,schleicher2011}. This function, however, significantly deviates from cometary dust observations at $\alpha\gtrsim170^\circ$, increasingly underestimating the amplitude of the brightness surge at higher $\alpha$ \citep{hui2013,knight2024}.

While micron-sized grains are typically optically dominant in both cometary and interstellar dust at low $\alpha$, the narrower but more intense diffraction peak of larger grains causes increasingly larger grains to become optically dominant at higher $\alpha$ where the two size distributions deviate. The greater relative abundance of larger grains in cometary dust \citep{fulle2004} supports a much steeper and stronger brightening at high $\alpha$ than interstellar dust, hence why a phase function model intended for the latter would underestimate the forward scattering brightness of the former.

We instead opted to use the Fournier--Forand phase function, which is an analytic approximation to Mie scattering derived from the anomalous diffraction approximation for an ensemble of particles with a power-law size distribution \citep{fournier1994}. This function was developed to model light propagation through ocean water with suspended particulates, which follow power-law size distributions much like cometary dust grains. Moreover, since forward scattering efficiency depends on the wavelengths of the scattered light relative to the size of the scattering grains, the nearly solar LASCO orange--clear color of both (139359) and 2P at high $\alpha$---despite the orange bandpass having a much shorter effective wavelength than the clear bandpass---supports the use of such a scale-invariant size distribution for their optically bright dust.

Most usefully, the Fournier--Forand model is parametrized by two physical parameters---the cumulative size index $\mu_d$ (defined such that the number of grains larger than a given radius $r_d$ is proportional to $r_d^{-\mu_d}$) and an effective optical refractive index $n_\mathrm{ref}$---which can be fitted to an observed forward scattering brightness surge to constrain the corresponding physical properties of the comet's dust. We used a functional form for the $\theta$-dependence of dust brightness $\Phi(\theta)$ derived from \citet{fournier1999}:

\begin{equation}
\begin{aligned}
\Phi(\theta)=&\frac{\gamma_1(\theta)\nu-\gamma_\nu(\theta)+[\gamma_\nu(\theta)\delta(\theta)-\gamma_1(\theta)]\sin^{-2}(\theta/2)}{4\pi\gamma_1(\theta)^2\delta(\theta)^\nu} \\
&+\frac{(1-\delta_{180}^\nu)(3\cos^2\theta-1)}{16\pi(\delta_{180}-1)\delta_{180}^\nu}
\end{aligned}
\end{equation}

\noindent where

\begin{equation}
\begin{aligned}
\nu&\equiv\frac{3-\mu_d}{2}\\
\delta_{180}&\equiv\frac{4}{3(n_\mathrm{ref}-1)^2}\\
\delta(\theta)&\equiv\delta_{180}\sin^2\left(\frac{\theta}{2}\right)\\
\gamma_k(\theta)&\equiv1-\delta(\theta)^k
\end{aligned}
\end{equation}

Note that $\mu_d$ represents the size index for dust grains within the photometric aperture, which is \emph{not} equal to the size index for dust grains produced by the comet. This distinction is because smaller grains---which accelerate more rapidly under solar radiation---spend less time within the aperture than larger grains, leading the observed dust population to be biased toward the slower, larger grains. For a simplistic correction of this effect, consider that the distance $\Delta L$ traveled by a grain accelerated from rest relative to the nucleus by radiation pressure at $a_\mathrm{rad}\propto\beta$ over time $\Delta t$ is $\Delta L=a_\mathrm{rad}\Delta t^2/2 \propto\beta\Delta t^2$. Now suppose the photometric aperture is set such that all grains $\Delta L<L_\mathrm{ap}$ fall within the aperture: the residence time $t_d$ within is then simply the time grains take to reach $\Delta L=L_\mathrm{ap}$, so $t_d=(2L_\mathrm{ap}/a_\mathrm{rad})^{1/2}\propto\beta^{-1/2}\propto r_d^{1/2}$. If dust production is considered relatively uniform over these timescales, the size index of dust grains actually produced by the comet gains an additional $\Delta\mu=1/2$ to become $\mu_d^*=\mu_d+\Delta\mu$.

However, at the highest $\alpha$ near inferior conjunction, radiation pressure accelerates the dust almost directly toward the observer, so dust flowing down the tail will not exit the aperture until either (1) its initial ejection velocity causes it to diffuse out, corresponding to an apparent coma larger than the photometric aperture, or (2) the dust is first removed by the physical curvature of the tail, a higher order effect requiring consideration of the comet's motion around the Sun. Case 1 does not apply to the presented data, where the dust has clearly been shaped into a tail before exiting the $3'$ radii apertures (see Figures~\ref{fig:img} and \ref{fig:syn}); i.e., the ejection speed satisfies $v_\mathrm{ej}t_d\ll L_\mathrm{ap}$, and we evaluate $t_d$ for case 2 under this limit. To do so, consider a reference frame fixed to the nucleus that rotates with the sunward direction at angular speed $\Omega_s$: over short time frames of a few days where $r$ and $\Omega_s$ are nearly constant, the forces acting on dust in this frame can be considered time invariant, so grains released at rest with the same $\beta$ all follow the same, overlapping trajectory matching the syndyne of that $\beta$, permitting analytic approximations for both the shapes of the syndyne and the age of dust grains along its length. As before, the antisunward acceleration is driven by $a_\mathrm{rad}$, but the rotation of this reference frame introduces a Coriolis force $a_\mathrm{cor}\propto\Delta\dot{L}\propto\beta\Delta t$ that deflects the dust toward the comet's negative heliocentric velocity direction by $\Delta Y\propto\beta\Delta t^3$. At very high $\alpha\approx180^\circ$, the photometric aperture cuts off dust at a nearly constant $\Delta Y$ (instead of $\Delta L$), leading to a slightly different residence time $t_d\propto\beta^{-1/3}$ (instead of $\beta^{-1/2}$) and a size index correction factor $\Delta\mu=1/3$ (instead of 1/2). Our observations span a range of $\alpha$ over which both cases may be applicable, so we opted to use a $\Delta\mu=0.4\pm0.1$ encompassing both correction factors.

Note also that this model does not directly describe the dust expected in cometary tails, as cometary dust grains are porous aggregates of submicron-sized monomers \citep{levasseurregourd2018} while the Fournier--Forand phase function models the scattering of effectively solid particles. High porosity grains can exhibit substantially different scattering behavior from more compact grains, with stronger forward scattering peaks reflective of their substructure \citep{debroy2017}. Fluffy grains with porosities far exceeding 0.9 have been found in the inner coma of comet 67P/Churyumov--Gerasimenko, but are only estimated to contribute $\lesssim$15\% of the dust cross section there \citep{fulle2015}. Moreover, porosity correlates inversely with tensile strength \citep{kimura2020,kreuzig2024}, and fragmentation mechanisms like solar radiation torque may be expected to preferentially destroy the more fragile, higher porosity grains well within the few days it would take the dust to traverse the photometric aperture \citep{herranen2020}. We consider the observed dust to be optically dominated by low ($\lesssim$0.5) porosity compact grains---consistent with the estimated porosity of Taurid meteoroids associated with 2P \citep{babadzhanov2009}---which can be reasonably approximated in forward scattering as solid grains of a porosity-weighted $n_\mathrm{ref}$, and thus be described there by the Fournier--Forand model.

In addition to dust, the photometry also captures emission by several different gaseous species, which does not dependent significantly on $\alpha$. The clear and orange filters used by all of our LASCO C3 and C2 photometry, respectively, are primarily sensitive to C$_2$ and NH$_2$, while HI1 is primarily sensitive to CN, C$_3$, and NH$_2$, with total gas emission contributing a comparable fraction of the total brightness through these filters for comets with typical gas ratios \citep{jones2018}. The true coma gas composition of (139359) is not known, so we approximate the dust-to-gas brightness ratio at each $\alpha$ as constant in all of our photometric bandpasses, and treat 2P similarly for simplicity. Both the formation and destruction length scales of these coma species are well below $3'$ for all considered LASCO and HI1 observations, so the full coma is effectively always contained within the photometric aperture, and we do not apply any aperture corrections to this contribution. Note that the LASCO observations cover the comets only at high $\alpha$ where the forward scattering dust overwhelms the gas emission, while C$_3$ and NH$_2$ emission are typically far weaker than CN emission within the HI1 sensitivity ranges \citep{ahearn1995,fink2009}, so the gas emission brightness measured from the combined dataset likely primarily reflects CN.

We quantified the level of forward scattering brightening seen by SOHO LASCO by first correcting all of the photometry for the observer--comet distance $\varDelta$, then subtracting the baseline magnitude interpolated from the backscattering ($\alpha<90^\circ$) STEREO-A HI1 photometry, which isolates the magnitude difference corresponding to forward scattering brightening. We used a free parameter $d_{90}\in[0,1]$ to represent the fractional dust contribution to the HI1 photometry at $\alpha=\theta=90^\circ$, and only applied tail correction and the Fournier--Forand phase function to this dust component of the coma brightness. Since the phase function is relatively flat at $\alpha\lesssim90^\circ$ ($\theta\gtrsim90^\circ$), $d_{90}$ closely approximates the fractional dust brightness contribution over much of this range as well. Note, however, that the Fournier--Forand model does not attempt to capture the highly structure-dependent scattering behavior toward $\alpha\sim0^\circ$, so should not be used be on its own if the dust contribution to backscattering is important (i.e., $d_{90}\gtrsim0.5$) and accurate behavior is needed at these low $\alpha$.

We constructed a likelihood function incorporating the LASCO--HI1 magnitude differences at their corresponding $\theta$, with $d_{90}$, $\mu_d$, $n_\mathrm{ref}$ as free parameters. We used uniform priors for $d_{90}$ and $\mu_d$, but found the model to be minimally sensitive to $n_\mathrm{ref}\sim1$--2 at the $\alpha>155^\circ$ ($\theta<25^\circ$) of our data, so set a prior of $(n_\mathrm{ref}-1)/n_\mathrm{ref}$ at $n_\mathrm{ref}>1$. This prior distribution gives a mean $\pm$ standard deviation of $1.3\pm0.2$ with a hard cutoff at $n_\mathrm{ref}=1$ and a long upper tail. We then performed Markov-chain Monte Carlo (MCMC) sampling of the resulting posterior distribution to constrain $d_{90}$ and $\mu_d$ using \texttt{emcee} \citep{foreman2013}.

\begin{figure*}
\centering
\includegraphics[width=0.53\linewidth]{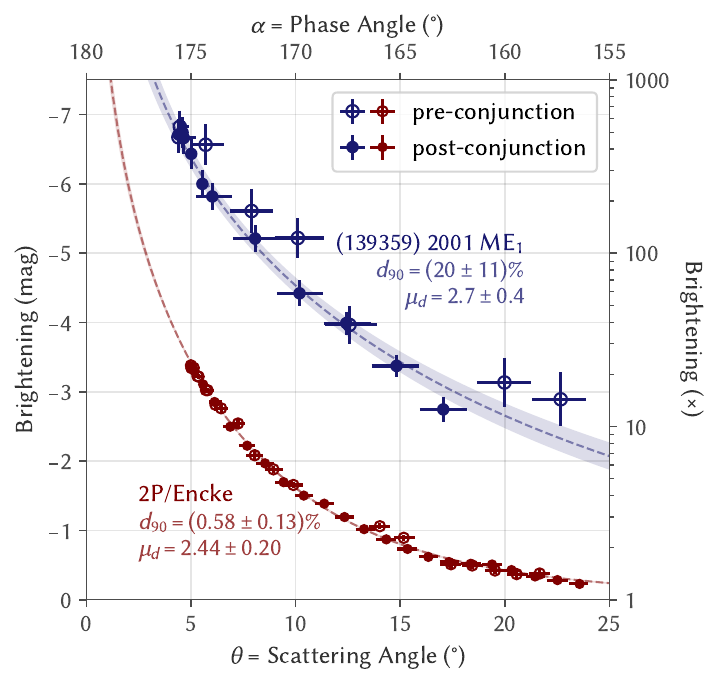}%
\includegraphics[width=0.46\linewidth]{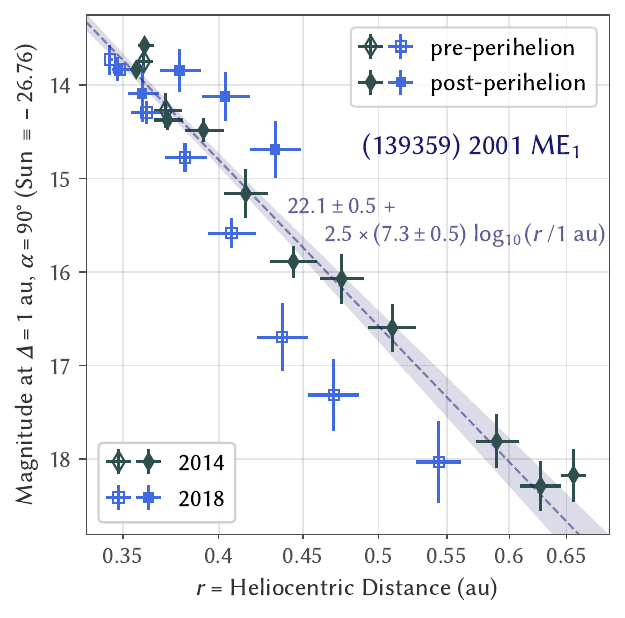}
\caption{Left: Forward scattering brightening over $\alpha=\theta=90^\circ$ measured from simultaneous high $\alpha$ (low $\theta$) SOHO LASCO observations and low $\alpha$ (high $\theta$) STEREO-A HI1 observations of (139359) 2001~ME$_1$ in 2018 compared with 2P/Encke in 2017, with fitted Fournier--Forand phase functions. The fitted $d_{90}$ are the fractional dust contributions to the brightness within the $3'\approx160{,}000$~km radius HI1 photometric apertures at $\alpha=\theta=90^\circ$, while $\mu_d$ are the cumulative dust size distribution power indices. Right: Variation of nucleus-subtracted and $\alpha$/$\varDelta$-corrected brightness of (139359) 2001~ME$_1$ with $r$ measured from STEREO-A HI1 observations in 2014 and 2018.}
\label{fig:forward}
\end{figure*}

Figure~\ref{fig:forward} (left) shows the LASCO--HI1 magnitude differences for both (139359) and 2P, adjusted to LASCO magnitudes above their $\alpha=\theta=90^\circ$ levels by our model, along with the fitted phase functions. The most prominent difference between the curves is the relative strength of (139359)'s forward scattering brightness surge compared with that of 2P. Even at the same $\alpha=175^\circ$, (139359)'s 6.4~mag brightening dwarfs 2P's 3.4~mag surge. This difference evidently arises from (139359)'s much higher $d_{90}=(20\pm11)\%$ compared to 2P's $d_{90}=(0.58\pm0.13)\%$, implying the former features a much larger dust contribution to its optical brightness than the latter.

The standard Schleicher--Marcus parametrization of the classic Henyey--Greenstein forward scattering model was fitted to and appears reasonably accurate for comets up to $\alpha\approx160^\circ$ \citep{marcus2007,schleicher2011}. The 2.7~mag surge of (139359) at $\alpha=160^\circ$ is also consistent with $d_{90}\approx20\%$ in that model, indicating the two both models agree at these lower $\alpha$ down to side-scattering geometry. The corresponding 0.4~mag brightening of 2P yields $d_{90}\approx0.9\%$, which is modestly above the $(0.58\pm0.13)\%$ of the Fournier--Forand fit. However, both models indicate that the brightnesses of the two comets are dominated by gas emission outside of forward scattering, which effectively suppresses the contribution of---and therefore the error arising from---the Fournier--Forand model in the overall (dust + gas) scattering function at $\alpha<90^\circ$.

Comet 2P is known to have an exceptionally low optical dust-to-gas brightness ratio from ground-based observations, with (139359)'s much higher $\delta_{90}\equiv d_{90}/(1+d_{90})=0.25\pm0.14$ actually falling much closer to average \citep{newburn1985,marcus2007}. Comets with lower $q$ tend to have lower optically visible dust relative to their gas emission than comets with higher $q$ \citep{ahearn1995}, suggesting 2P's anomalously low $d_{90}$ may have been acquired through thermal evolution. While 2P's $q=0.34$~au is nearly identical to that of (139359) at present, the former's somewhat smaller aphelion distance of $Q=4.1$~au compared to the latter's $Q=4.9$~au keeps 2P farther from Jupiter on a more dynamically stable orbit \citep{bottke2002} that may have kept that comet closer to the Sun for longer. We discuss the dynamical history of (139359) further in section~\ref{sec:dynamics}.

The fitted cumulative size indices $\mu_d^*=3.1\pm0.4$ for (139359) and $2.8\pm0.2$ for 2P are statistically indistinguishable from each other, and represent the size indices for the optically dominant micron-sized dust. For comparison, the latter measurement is modestly above the $\mu_d^*=2.4\pm0.2$ previously determined from nighttime observations of 2P through modeling the backscattering and thermal emission morphology of the millimeter-sized dust \citep{sarugaku2015}. Both of our values are also comparable with the higher end of the $\mu_d^*\sim2$--3 typical of comets modeled from similar nighttime observations \citep{fulle2004}. Steeper-than-average size distributions over the micron size range might be at least partially attributable to the low $r$ of the observations and associated enhancement of grain fragmentation mechanisms like radiative torque \citep{herranen2020}, as discussed earlier.

\subsubsection{Activity Level}
\label{sec:activity}

To evaluate variations in (139359)'s level of cometary activity, we first used the brightness--phase model for (139359) from the previous section to correct the HI1 photometry from both the 2014 and 2018 apparitions to remove the impact of forward scattering. In practice, only a few of the HI1 observations near the end of the 2014 apparition and the beginning of the 2018 apparition were significantly affected by forward scattering, given the optical dominance of gas emission at most $\alpha$.

Figure~\ref{fig:forward} (right) shows the comet's resulting $\varDelta$- and $\alpha$-normalized magnitudes as a function of $r$ from both apparitions. The photometry shows the comet somewhat brighter at the same distance post-perihelion than pre-perihelion during the 2018 apparition, although we caution the faintness of the comet near the limit of HI1 sensitivity leaves the photometry prone to slowly varying systematic offsets of this order from coronal fluctuations, which affects background determination. A power-law least-squares fit to the combined data set finds the comet's brightness proportional to $r^{-7.3\pm0.5}$, with an $r=1$~au extrapolated magnitude of $22.1\pm0.5$ much fainter than the nucleus itself at low $\alpha$. The prior nondetections of cometary activity at $r\gtrsim1$~au therefore remain consistent with the extrapolated brightness fall-off. The model also places the comet at magnitude 12 during its 2001 LASCO C3 transit, comparable to the limiting magnitude of that data and consistent with its lack of a clear detection.

The fitted peak magnitude of $13.51\pm0.08$ at $r=q=0.34$~au and low $\alpha$ is 5.2~mag fainter than measured for 2P at perihelion in 2017, and 5.8~mag fainter than 2P's peak brightness 2--4~days later. After removing the $-2.5\log_{10}(1-d_{90})\approx0.2$~mag dust contribution to (139359)'s brightness and the negligible dust contribution to 2P's brightness, (139359)'s gas emission near perihelion is therefore 5.4~mag below 2P's gas emission near perihelion, and 6.0~mag below it 2--4~days later, all at nearly the same $r=0.34$~au. Lyman-$\alpha$ observations of 2P indicate its water production rate to be ${\sim}5\times10^{28}$~molecules~s$^{-1}$ near perihelion and ${\sim}7\times10^{28}$~molecules~s$^{-1}$ around 2~days later \citep{raymond2002}. If (139359) has a similar gas coma composition as the fairly typical coma composition of 2P \citep{ahearn1995}, the gas production rates of the two comets near perihelion should be nearly proportional to the total brightness of the coma gas emission there. 

Using this scaling, (139359) has a water production rate of ${\sim}3\times10^{26}$~molecules~s$^{-1}$ near perihelion. For comparison, a similarly sized $\sim$3--4~km diameter nucleus fully covered with near-surface water ice in contact with dark material would be sublimating near ${\sim}1\times10^{30}$~molecules~s$^{-1}$ \citep{prialnik2004,rivkin2005,masiero2021}---comparable to the rate of the fresh Oort cloud comet C/2011~L4 (PANSTARRS) at this $r\sim0.3$~au \citep{combi2014}. The water production rate derived for (139359) corresponds to an active fraction of just $\sim$0.02--0.03\%, which is among the lowest identified for any near-Earth comet---comparable to those of the weakly active comets 209P/LINEAR \citep{schleicher2016} and 460P/PANSTARRS \citep{li2017}---and suggests that (139359) has little remaining near-surface ice available to drive activity.

Since the observed brightness of (139359) outside forward scattering is predominantly from gas emission, and gas lifetimes scale as $r^2$ while their fluorescence efficiencies scale as $r^{-2}$, the production rates of the emitting gases roughly scale with the same $r$-dependence as overall brightness. Then, if gas ratios remain similar throughout the comet's activity, the presumably predominant water outgassing should likewise fall off away from perihelion as $r^{-7.3\pm0.5}$. This outgassing dependence that is much steeper than than $r^{-2}$ suggests that water may not be efficiently sublimating with respect to the incident sunlight on the nucleus, despite the comet's proximity to the Sun. This behavior, which is common among Jupiter-family comets \citep{ahearn1995,combi2019}, could potentially be attributed to the escaping water vapor originating from hydrated compounds more refractory than water ice \citep{mason1963} or the sublimating ice being largely insulated by an overlaying mantle of refractory material \citep{kuhrt1994}, with both options plausible for a thermally evolved nucleus that has lost most of its near-surface ice.

Meanwhile, (139359)'s dust is 2.8~mag fainter than that of 2P near perihelion at $\alpha=175^\circ$. \citet{sarugaku2015} derived 2P's dust production rate at perihelion during its 2003 apparition to be ${\sim}10^2$~kg~s$^{-1}$ with a maximum dust radius $r_d$ of 4--10~cm.\footnote{\citet{sarugaku2015} actually reported a peak dust production rate of ${\sim}10^4$~kg~s$^{-1}$ in an outburst by the comet one week after perihelion---perhaps similar to one recorded by LASCO during the previous apparition in 2000 \citep{lamy2003}---but we see no evidence for such an outburst in our 2017 data, so disregard this outburst-enhanced rate for our analysis. An earlier study by \citet{epifani2001} estimated a peak of $2\times10^3$~kg~s$^{-1}$ two weeks before the 1997 perihelion, which could conceivably have been affected by a similar outburst unresolved by their data/modeling.} Next, if the two comets had identical dust size distributions, the dust production rate of (139359) relative to that of 2P near perihelion would scale proportionally to their dust brightness under their comparable viewing geometries. However, while the dust size indices $\mu_d$ of the two comets were indistinguishably close in our forward scattering analysis, (139359)'s far weaker gas production may not be capable of ejecting grains as large as 2P produces.

Assuming (139359) has a nucleus of bulk density $\sim$0.5~g~cm$^{-3}$, dust grains of bulk density $\rho_d\sim1$~g~cm$^{-3}$ with drag coefficients on the order of $\sim$1, and water outgassing from the surface near the blackbody thermal speed (i.e., net $\sim$400~m~s$^{-1}$ outflow) broadly distributed over the sunward side of the nucleus, the force of gravity exceeds that of gas drag on the surface of the nucleus for dust grains larger than about $r_d\sim1$~mm near perihelion. With this smaller maximum $r_d$, a minimum $r_d\ll1$~mm, and a canonical $\mu_d\sim2.5$ in the millimeter size range, we estimate through scaling from 2P's (micron-sized) dust brightness that the perihelion dust production rate is on the order of $\sim$1~kg~s$^{-1}$, corresponding to a dust-to-gas mass loss ratio on the order of $\sim$0.1.

Note that outgassing need not be perfectly uniform over the surface for the above approximation to be valid; in fact, interparticle cohesion would hold back all grains on (139359)'s nucleus in this case \citep{gundlach2015}. Non-collimated outgassing from small active areas spread across the nucleus can produce the locally higher pressures needed to dislodge grains. Once above the surface, small-scale pressure variations smooth out and the gravity--gas drag balance becomes roughly as discussed. However, the maximum $r_d$ and therefore dust production rate could be much higher if (139359)'s outgassing is concentrated within a few collimated jets---thus elevating gas pressure even well above the surface---or if (139359) is in an excited spin state close to its gravitational binding limit---thus reducing the gas pressure needed to eject dust grains of any given size.

In principle, it is also possible to derive dust production rates directly from just our forward scattering dust photometry in conjunction with an appropriate dust scattering model and assumed size distribution, without scaling from a known rate. For very rough estimates, consider a crude scattering model of dust grains as opaque disks that diffract light into Airy disk patterns. We calculated the scattering efficiency---and thus quantity---of dust captured by the LASCO C2 forward scattering observations of both comets at maximum $\alpha$ for grain sizes distribution with the $\mu_d$ fitted above (i.e., ignoring potential breaks in the size distribution), a minimum $r_d=1$~$\mu$m, and maximum $r_d=7$~cm for 2P (the midpoint of \citealt{sarugaku2015}'s 4--10~cm range) and 1~mm for (139359). We then estimated the residence time within the photometric aperture for grains of each $r_d$ and found dust production rates on the orders of ${\sim}10^2$--$10^3$~kg~s$^{-1}$ for 2P and $\sim$1--10~kg~s$^{-1}$ for (139359), which are, indeed, consistent with the earlier, scaled estimates.

\subsection{Dynamical Evolution}
\label{sec:dynamics}

We investigated the recent thermal history of (139359) through the dynamical evolution of its trajectory. We used \texttt{REBOUND} \citep{rein2012} with its standard IAS15 integrator \citep{rein2015} to numerically integrate the trajectories of 100 orbital clones drawn from the orbital elements and associated covariance matrix of JPL orbit \#130 for 30~kyr into the past and future. Our simulation included the gravity of the Sun and the eight major planets (combined with the masses of their moons at their barycenters) with post-Newtonian relativistic corrections. The simulation, however, neglected the outgassing-driven acceleration of the comet, which is not easily modeled far into the past or future, and were excluded entirely in the derivation of this orbital solution.

\begin{figure}
\centering
\includegraphics[width=\linewidth]{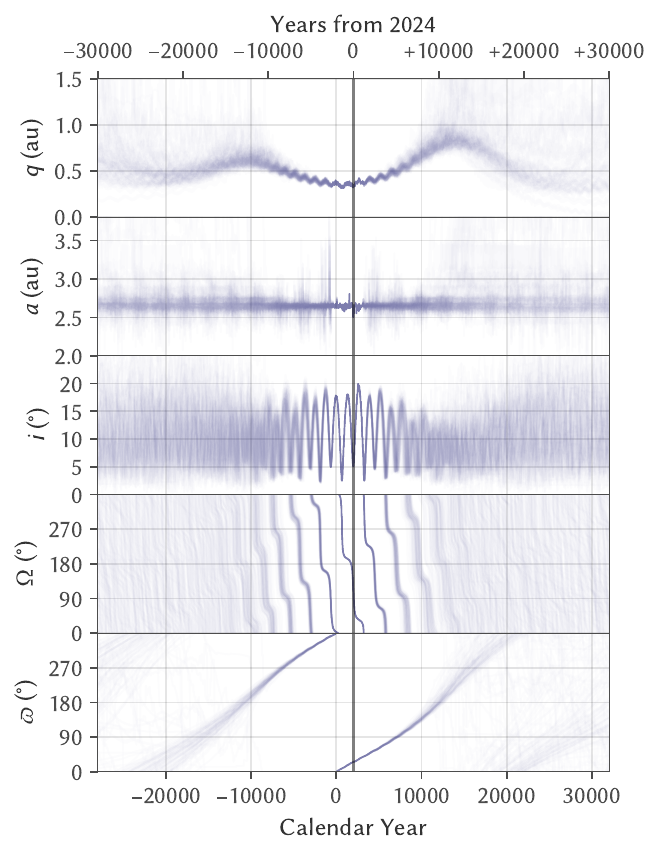}
\caption{Evolution of perihelion distance $q$, semi-major axis $a$, inclination $i$, longitude of ascending node $\Omega$, and longitude of perihelion $\varpi$ of 100 orbital clones of (139359) 2001~ME$_1$ for 30~kyr into the past and future, demonstrating the stability of its near-Earth orbit over $\sim$10~kyr timescales.}
\label{fig:evolve}
\end{figure}

Figure~\ref{fig:evolve} shows the resulting evolution of the comet's osculating orbital elements over the $\pm$30~kyr integration period. All clones remained gravitationally bound to the solar system for the duration of the integration period, except for a single clone (1\% of all clones) in the forward integration. Most clones also remained near-Earth comets for the duration for the period, with 77\% in the reverse integration and 58\% in the forward integration retaining a perihelion distance of $q<1.3$~au and a semi-major axis of $a<34$~au (orbital period $P<200$~yr).

These results indicate (139359) has likely followed and thus been subjected to the solar heating of a near-Earth orbit for the past ${\sim}10^4$--$10^5$~yr, which is toward the upper end of Jupiter-family comet dynamical lifetimes \citep{fernandez2014}. Given that (139359) also has, and has had a lower $q$ and thus greater intensity of solar heating near perihelion than most Jupiter-family comets for much of this period, its unusually low level activity likely resulted from the depletion of near-surface ices through sublimation during this long period with intense perihelic heating.

\subsection{Meteor Activity}

Millimeter-sized and larger dust grains ejected from a comet nucleus over repeated perihelion passages initially spread out in a dense trail along the orbit of the comet, before being eventually dispersed across the solar system by planetary perturbations and various nongravitational forces \citep{jenniskens1998}. Consequently, comets with orbits approaching Earth's orbit often produce meteor showers when Earth crosses through the associated dust trail each year. Meteor observations can provide a window into the past activity of these comets.

\begin{figure}
\centering
\includegraphics[width=\linewidth]{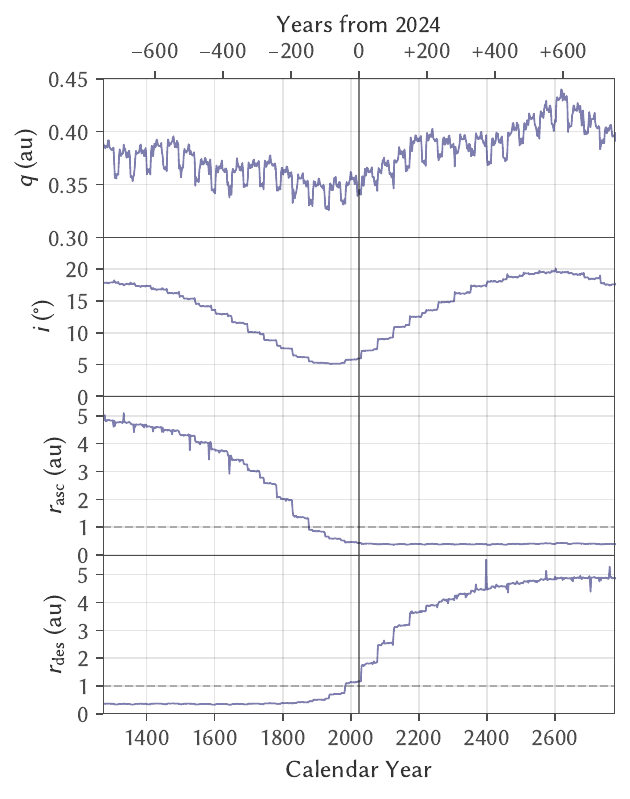}
\caption{Near-term evolution of (139359) 2001~ME$_1$'s perihelion distance $q$, inclination $i$, ascending node distance $r_\mathrm{asc}$, and descending node distance $r_\mathrm{des}$ over the previous and next 750~yr. The orbit is most conducive to meteor showers on Earth when $r_\mathrm{asc}\sim1$~au or $r_\mathrm{des}\sim1$~au, corresponding to a near-intersection of the comet's orbit with that of Earth's near the respective node.}
\label{fig:evolve_short}
\end{figure}

The orbit of (139359) presently passes very near the orbit of Earth near the former's descending node, with a minimum orbit intersection distance (MOID) of only 0.013~au in 2024. In fact, Figure~\ref{fig:evolve_short} shows the descending node of the comet's osculating orbit only recently crossed over Earth's orbit in 1982, and will rapidly drift outward over the next few centuries. Meteor activity should therefore have been at an elevated level in recent decades, becoming less favorable in the next few decades as continued perturbations gradually shift the comet's orbit and dust trail away from Earth.

Despite this favorable geometry, the June $\xi^1$-Sagittariids meteors associated with (139359) have only recently been reported. \citet{ye2016a} investigated possible material ejected from this object and identified a previously unreported meteor outburst in 2006 CMOR radar data. Later, \citet{jenniskens2023} analyzed video data collected over 2010--2023 by the Cameras for Allsky Meteor Surveillance (CAMS) and reported only extremely weak activity, with an average zenithal hourly rate (ZHR, a measure of meteor flux; \citealt{koschack1990}) of 0.04. A simple meteoroid model by \citet{ye2016a} based on blind assumptions of the comet's then-unseen activity suggested that the 2006 meteor outburst originated from dust ejected between 1924 and 1967.

We reevaluated the origin of the 2006 meteor outburst in light of the direct observations of (139359)'s cometary activity and investigated whether similar outbursts occurred in other years. \citet{ye2016a} estimated that only dust ejected within the last 800~yr still remain in sufficiently concentrated streams to plausibly produce observable meteor activity traceable back to this comet. We therefore adopted this same cutoff and simulated the dust production activity of (139359) over the last 800~yr using the approach described in \citet{ye2016b}.

We integrated the comet's orbital motion starting from JPL orbit \#130 (epoch 2024) and proceeded backward 800~yrs, releasing meteoroids when the comet was within 1~au of the Sun. The simulation used a Bulirsch--Stoer integrator \citep{stoer1980} and included the Sun and the eight major planets (combined with their moons at their barycenters) as gravitational masses, as well as the comet and meteoroids as test particles.

All meteoroids were integrated forward until year 2050. Simulated meteoroids spanned radii of $r_d=0.3$--3~mm and had a bulk density of $\rho_d=1$~g~cm$^{-3}$, consistent with the nominal value used in earlier calculations. The smallest grains of $r_d=0.3$~mm roughly corresponds to the minimum size detectable with CMOR \citep{jones2005}, while the largest $r_d=3$~mm grains are on the upper end of our $r_d\sim1$~mm order-of-magnitude estimate for the maximum grain size liftable by the comet's modern-day level of activity derived in section~\ref{sec:activity}. Dust production followed the comet's nominal modern-day rate and $r$-dependence, as $(2~\mathrm{kg~s^{-1}})\times(r/0.34~\mathrm{au})^{-7.3}$ as derived in section~\ref{sec:activity} and scaled to $r_d<3$~mm. Ejected grain sizes followed a cumulative power-law distribution with a  negative index of $\mu_d^*=2.5$, as typical for meteoroid streams \citep[e.g.,][]{blaauw2011}.

We set the initial velocity of the ejected dust grains as the terminal velocity beyond the influence of gas drag and the nucleus' gravity. Ejection directions were randomly distributed over the sunward hemisphere, although tests with an isotropic ejection model indicate the hemispheric restriction did not substantially affect the final results.

The terminal ejection speed $v_\mathrm{ej}$ from a nucleus with abundant near-surface water ice is well represented by the gas drag model of \citet{whipple1950}. For (139359) at perihelion, this Whipple speed is $v_W\sim60$--70~m~s$^{-1}$ for $r_d=1$~mm grains, with an approximate scaling relation of $v_W\propto r_d^{-1/2}$. However, as found in section~\ref{sec:activity}, the outgassing rate at perihelion is only $\sim$0.02--0.03\% that of an efficiently sublimating nucleus. Assuming the activity that is present is evenly spread over the sunlit hemisphere, the reduction in outgassing corresponds to a reduction in the gas density to $\sim$0.02--0.03\% that of the Whipple model. Since the terminal speed (when far below the gas outflow speed of a few km~s$^{-1}$; \citealt{combi2004}) scales with the square root of gas density, $v_\mathrm{ej}$ becomes $\sim$1--2\% of $v_W$.

Note that the actual $v_\mathrm{ej}$ may be somewhat higher than for this uniformly distributed activity model if all of the outgassing is instead concentrated in a few active regions covering a small fraction of the surface. Concentrated activity would produce higher gas densities and thus greater acceleration of dust from those active regions. For example, observations of the similarly active 209P indicated that comet had a $v_\mathrm{ej}$ that was several times higher, at $\sim$10\% of $v_W$ \citep{ye2016b}.

We can also directly constrain $v_\mathrm{ej}$ from the morphology of the coma and inner tail, since higher $v_\mathrm{ej}$ will expand the size of the coma and width of the inner tail. Figure~\ref{fig:syn} shows the coma to be unresolved, but the tail to be ${\sim}1'$ across at ${\sim}1'$ from the nucleus. Therefore, in the $\sim$3~days it took for the optically bright $\beta\sim0.1$ dust ejected at zero velocity to drift tailward by ${\lesssim}1'$, the envelope of dust ejected at that same time had expanded to a width of ${\sim}1'$, or a projected $\sim$13,000~km from the zero-velocity syndyne, at $\sim$50~m~s$^{-1}$. However, the dust has a non-zero range of $\beta$ that also contributes to dispersion in this orthogonal direction, and the measured width is not much wider than the instrumental point-spread function, so we consider this speed to only be an upper limit for the actual $v_\mathrm{ej}$ of these $r_d\sim3$~$\mu$m grains. Equivalently, $v_\mathrm{ej}\lesssim3$~m~s$^{-1}$ for $r_d\sim1$~mm grains, or $\lesssim$5\% of the $v_W\sim60$~m~s$^{-1}$ at the $r=0.37$~au of the observation---suggesting the true value may be quite near the 1--2\% of $v_W$ expected from uniformly distributed outgassing. In lieu of tighter constraints, we used $v_\mathrm{ej}=0.02v_W$ for our simulation.

At the end of the simulation, we computed the meteoroid flux at Earth based on the number of simulated particles in Earth's vicinity. Note that we previously set our simulated minimum grain radius of $r_d=0.3$~mm to match the radar sensitivity of CMOR, and many of the smaller simulated meteors would not be detectable visually or by a video system like CAMS. In fact, CAMS has a limiting $V$ magnitude of around +4 \citep{jenniskens2011}, which corresponds to $r_d=1.5$~mm for meteoroids of typical composition with our assumed 1~g~cm$^{-3}$ density at their incident 30~km~s$^{-1}$ \citep{campbell2004}.

By convention, ZHR is defined for an ideal visual limiting magnitude of +6.5, corresponding to a minimum $r_d=0.5$~mm which real observations of meteor flux must be corrected to in order to calculate the ZHR \citep{koschack1990}. Real observations of weak meteor outbursts, as we are considering, record too few meteors to accurately determine the true size distribution, and typically use an assumed power-law cumulative radius distribution with index $-2.5$ to scale the recorded meteor flux for ZHR calculations. This size distribution is identical to the one we assumed for dust grains ejected from the comet (i.e., $\mu_d^*=2.5$), but grains become spatially sorted by size pressure through solar radiation. An individual meteor outburst can feature just a narrow range of grain sizes---very different from the assumed power-law distribution---leading to enormous differences in the ZHRs derived by observations with different sensitivity limits. The 2012 Draconids outburst, for example, produced radar-derived ZHRs ${\sim}50\times$ that of the ZHRs derived from visual observations due to this effect \citep{ye2014}. We consider both video observations from CAMS and radar observations from CMOR, so we followed the same approach to derive separate ZHRs for a video regime limited to $r_d>1.5$~mm, corresponding to the former, and a radar regime of $r_d>0.3$~mm (i.e., all simulated meteors), corresponding to the latter.

Due to the large uncertainty in the derived maximum grain size $r_d\sim1$~mm ejectable by the comet, we also computed separate ZHRs for the nominal $r_d<1$~mm and for $r_d<3$~mm (i.e., all simulated grains) half an order of magnitude higher (the half-order-of-magnitude lower constraint of $r<0.3$~mm produces no meteors detectable with either CAMS or CMOR). Still, given the considerable uncertainties in the comet's dust ejection velocity, production rate, and size distribution, we consider derived meteoroid fluxes to be order-of-magnitude estimates. The encounter times depend almost entirely on well-constrained planetary dynamics, so are usually accurate to within an hour \citep{egal2020}.

Table~\ref{tab:meteors} shows all of the meteor outbursts above the sporadic background level since the start of standardized meteor observations in the 1980s \citep{roggemans1988}. \citet{ye2016a}'s CMOR detection of a meteor outburst in 2006---the only detection in their survey between 2002 and 2015---evidently indicates the presence of $r_d>1$~mm grains, as the simulation expects no significant outburst otherwise. They estimated that outburst to have had a flux on the order of $\sim$0.01~km$^{-2}$~hr$^{-1}$, or ZHR $\sim10$. A reevaluation of the 2006 CMOR data, courtesy of P. Brown, could not significantly improve this flux estimate beyond establishing that the outburst was modestly weaker than the nearby antihelion sporadic source. The antihelion meteors produce a radar-equivalent ZHR of $\sim$20--30 \citep{brown1995}. We therefore estimate that the observed 2006 outburst had a radar-equivalent ZHR of $\sim$10--20.

\begin{deluxetable*}{cccccccccccccc}
\tablecaption{Modeled meteor outbursts from (139359) between 1980 and 2050}
\label{tab:meteors}

\tabletypesize{\small}
\tablecolumns{12}
\tablehead{
\colhead{Date (UT)} & Ejected\tablenotemark{a} & & \multicolumn{4}{c}{Video regime (CAMS; $r_d>1.5$~mm)} & & \multicolumn{4}{c}{Radar regime (CMOR; $r_d>0.3$~mm)} \\
\cline{4-7} \cline{9-12} 
& & & \colhead{Peak (UT)\tablenotemark{b}} & \colhead{FWHM\tablenotemark{b}} & \colhead{ZHR\tablenotemark{b,c}} & \colhead{Obs.\tablenotemark{d}} & & \colhead{Peak (UT)\tablenotemark{b}} & \colhead{FWHM\tablenotemark{b}} & \colhead{ZHR\tablenotemark{b,c}} & \colhead{Obs.\tablenotemark{d}}
}

\startdata
\multicolumn{12}{c}{Ejected dust size limit: $r_d<1$~mm} \\
\hline
1992 Jun 23 & 1898 & & \nodata & \nodata & \nodata & \nodata & & 13:03 & 1~hr & 18 & \nodata \\
2015 Jun 24 & 1971 & & \nodata & \nodata & \nodata & \nodata & & 19:39 & 0.7~hr & 15 & \nodata \\
\hline
\multicolumn{12}{c}{Ejected dust size limit: $r_d<3$~mm} \\
\hline
1981 Jun 24 & 1928 & & \nodata & \nodata & $<$2 & \nodata & & 16:28 & 0.2~hr & 32 & \nodata \\
1984 Jun 23 & 1924 & & 11:52 & 0.1~hr & 4 & \nodata & & \nodata & \nodata & $<$10 & \nodata \\
1992 Jun 23 & 1898 & & \nodata & \nodata & $<$2 & \nodata & & 13:26 & 4~hr & 29 & \nodata \\
1993 Jun 23 & 1954 & & \nodata & \nodata & $<$2 & \nodata & & 11:47 & 0.6~hr & 20 & \nodata \\
2006 Jun 23 & 1958 & & \nodata & \nodata & $<$2 & \nodata & & 23:54 & 0.6~hr & 19 & \nodata \\
2006 Jun 24 & 1962 & & \nodata & \nodata & $<$2 & \nodata & & 06:23 & 0.3~hr & 10 & $\sim$10--20 \\
2015 Jun 23 & 1958 & & \nodata & \nodata & $<$2 & $<$2 & & 05:07 & 5~hr & 12 & \nodata \\
2015 Jun 24 & 1971 & & \nodata & \nodata & $<$2 & $<$2 & & 17:51 & 2~hr & 27 & \nodata \\
\enddata

\tablenotetext{a}{Year in which the meteoroids contributing to each outburst was ejected from the comet.}
\tablenotetext{b}{Peak time, FWHM, and peak ZHR of the best fit Gaussian function to the modeled ZHR over time of each outburst. Modeled ZHRs are derived from the total total meteor flux of $r_d>1.5$~mm (video) and $r_d>0.3$~mm (radar) grains scaled to the standard ZHR limiting $V$ magnitude of 6.5 ($r_d>0.5$~mm) for a cumulative size distribution with a power index of $-2.5$, mirroring the derivation of ZHRs from real video/radar observations when the true size distribution is poorly constrained.}
\tablenotetext{c}{Upper limits are given for outbursts with peak ZHRs below the sporadic background near the radiant ($<$2 for video/CAMS and $<$10 for radar/CMOR; P. Jenniskens and P. Brown, private communications).}
\tablenotetext{d}{Observed ZHRs and upper limits, as discussed in the main text (\citealt{ye2016a}; P. Jenniskens and P. Brown, private communications).}
\end{deluxetable*}

The meteor shower radiant was only above the horizon for CMOR on 2006 June 24 from approximately 01:00 to 11:00 UT, during which only 1962 ejecta significantly contributed to meteor activity. The CMOR-derived ZHR of $\sim$10--20 is comparable to the corresponding simulated rate expected for an ejection size limit of $r_d<3$~mm, encompassing the $r=1$--1.5~mm grains of this outburst. Since the simulated ZHRs were derived from a dust production model compatible with the observed modern-day activity, the comet may have been active at a similarly low level in 1962 as at present.

Curiously, our models also show a similarly strong and well-timed outburst in 2015 on June 23, yet none was recorded by CMOR. However, a closer inspection of that data revealed that CMOR was hampered by a thunderstorm during the outburst time frame, over which almost no meteors were detected from any source (P. Brown, private communications), so this nondetection does not place a useful constraint on the strength of the outburst.

CAMS was active during both modeled outbursts in 2015, but recorded no associated meteors, setting limits of ZHR $<2$ (P. Jenniskens, private communications). However, these nondetections remain compatible with the simulation results as the contributing meteoroids were all of $r_d\approx1$~mm, falling just below the $r_d>1.5$~mm sensitivity range of CAMS.

Our simulation indicates there to be no additional significant meteor outbursts through the end of the integration in 2050. The descending node of (139359)'s orbit is now gradually drifting exterior to Earth's orbit, which will suppress future meteor activity for the next few centuries.

Note that the ascending node also crossed Earth's orbit in 1875, producing another period of time conducive to meteor activity in the late 19$^\text{th}$ century. However, the ascending node was on the outbound leg of the orbit during this period, leading any associated meteors to fall in a largely daytime meteor shower. Between the poor observability of these meteors and the general lack of detailed meteor records during this time period, we consider the absence of observation reports of meteors from (139359) during this earlier dust trail crossing unlikely to provide useful constraints on the comet's activity, and chose not to conduct a more detailed analysis of this event.

\section{Conclusions}

The minor planet and NEO (139359) 2001~ME$_1$ actually remains a visibly active comet, producing gas and dust near perihelion while unseen by classical nighttime observations. Favorable forward scattering geometry amplified the brightness of the comet's otherwise weak activity to become visible to SOHO's LASCO coronagraphs. We analyzed these forward scattering observations together with simultaneous backscattering observations by STEREO-A's HI1 camera, and found the following results:

\begin{enumerate}
\item The $\sim$7~mag forward scattering brightness surge of (139359) at phase angles up to $175^\circ\llap{.}6$ appears to be well fit by a Fournier--Forand scattering function for a cumulative particle size distribution with a power index of $-2.7\pm0.4$.
\item Corrected for the speed difference between differently sized grains in the tail, the corresponding size distribution of dust ejected from the nucleus has an index of $-3.1\pm0.4$ in the micron size range.
\item Dust contributes about $(20\pm11)\%$ of the side-scattering flux within the $3'\sim120{,}000$~km radius HI1 photometric apertures near perihelion. This dust was much brighter relative to the gas coma than for 2P/Encke, whose dust contributes only $(0.58\pm0.13)\%$ of that comet's side-scattering flux and produced only a $\sim$3~mag brightening under similar forward scattering geometry.
\item Assuming the coma has a typical gas composition, the water production rate peaks at ${\sim}3\times10^{26}$~molecules~s$^{-1}$ near perihelion, only $\sim$0.02--0.03\% that of an efficiently sublimating icy nucleus of the same 3--4~km diameter.
\item Cometary activity both rises and falls rapidly as the comet approaches and recedes from perihelion, with its side-scattering coma brightness---which is predominantly of gas emission---scaling with heliocentric distance as a steep power-law of index $-7.3\pm0.5$.
\item The coma and tail morphology near the forward scattering peak indicates the optically brightest $\beta\sim0.1$ (few micron-sized) grains seen there are ejected from the nucleus at $\lesssim$50~m~s$^{-1}$, or $\lesssim$5\% of the Whipple ejection speed expected for an efficiently sublimating icy nucleus, when scaled to grains in the millimeter size range.
\end{enumerate}

We also explored the dynamical evolution of (139359) through orbital integrations, and found the following:

\begin{enumerate}
\item The comet has likely followed a near-Earth orbit similar to its present one for over 10~kyr, and will likely continue to do so for at least another 10~kyr.
\item The descending node of the comet's orbit crossed from interior to exterior of Earth's orbit in 1982, which has supported meteor activity from the comet over the last few decades.
\end{enumerate}

Finally, we evaluated the meteor activity from the comet by performing an orbital simulation of meteoroids ejected over the past few centuries. Using cometary activity parameters consistent with values derived from the recent SOHO/STEREO observations, the simulation closely reproduced the CMOR observations of a meteor outburst in 2006 and found the responsible meteoroids to have had radii of $\sim$1--1.5~mm. This result indicates that the comet was also active in 1962, likely at a similarly low level as at present.

Collectively, these results appear to describe an aging Jupiter-family comet whose surface has become largely depleted of accessible ice after thousands of years of heating in the inner solar system. What ice remains is evidently sufficiently insulated that sublimation activity is effectively suppressed beyond Earth's orbit, visibly persisting only under the far more intense heating near perihelion. Targeted twilight observations of this comet may improve the characterization of its activity and help to better relate the properties of the comet to those of the many, more distantly active comets previously observed and studied in far greater detail. Future observations of other seemingly dormant comet nuclei on similarly closely Sun-approaching orbits may also uncover additional cases of residual activity near perihelion that could offer further insight into the mechanisms behind the behavior and evolution of such objects.

\begin{acknowledgments}
We thank Peter Brown (UWO) for help with reevaluating the CMOR data, and Peter Jenniskens (SETI) for revisiting the CAMS data. We also thank the three anonymous reviewers for their comments and suggestions.

Q.Z. was supported by a Percival Lowell Postdoctoral Fellowship from Lowell Observatory. Q.Y. and M.M.K. were supported by NASA program 80NSSC22K0772. K.B. was supported by the NASA-funded Sungrazer project. 

The SOHO/LASCO data are produced by a consortium of the NRL (USA), MPS (Germany), Laboratoire d'Astronomie (France), and the University of Birmingham (UK). SOHO is a project of international cooperation between ESA and NASA.

The STEREO/SECCHI data are produced by an international consortium of the NRL, LMSAL, NASA GSFC (USA), RAL and the University of Birmingham (UK), MPS (Germany), CSL (Belgium), and IOTA and IAS (France).

Based on observations obtained with the Samuel Oschin 48-inch Telescope at the Palomar Observatory as part of the Zwicky Transient Facility project. ZTF is supported by the National Science Foundation under Grant No. AST-1440341 and a collaboration including Caltech, IPAC, the Weizmann Institute for Science, the Oskar Klein Center at Stockholm University, the University of Maryland, the University of Washington, Deutsches Elektronen-Synchrotron and Humboldt University, Los Alamos National Laboratories, the TANGO Consortium of Taiwan, the University of Wisconsin at Milwaukee, and Lawrence Berkeley National Laboratories. Operations are conducted by COO, IPAC, and UW. This research has made use of the NASA/IPAC Infrared Science Archive, which is funded by the National Aeronautics and Space Administration and operated by the California Institute of Technology.

This research has made use of data and/or services provided by the International Astronomical Union's Minor Planet Center, and by the Jet Propulsion Laboratory's Small Body Database.

The views expressed in this article are those of the authors and do not reflect the official policy or position of the U.S. Naval Academy, Department of the Navy, the Department of Defense, or the U.S. Government.
\end{acknowledgments}

\facilities{SOHO (LASCO), STEREO (HI1), PO:1.2m}
\software{\texttt{Astropy} \citep{astropy2013,astropy2018,astropy2022}, \texttt{Astroquery} \citep{ginsburg2019}, \texttt{emcee} \citep{foreman2013}, \texttt{Matplotlib} \citep{hunter2007}, \texttt{NumPy} \citep{vanderwalt2011,harris2020}, \texttt{REBOUND} \citep{rein2012}, \texttt{sbpy} \citep{mommert2019},  \texttt{SciPy} \citep{virtanen2020}}

\bibliography{ms}
\bibliographystyle{aasjournalv7}

\end{CJK*}
\end{document}